\documentclass{article}
\usepackage{graphicx}
\usepackage{amsmath}

\input{tcilatex}

\begin{document}

{\Large On the Security of New Key Exchange Protocols Based on the Triple
Decomposition Problem}\medskip

\ \ \ \ \ \ \ \ \ \ \ \ \ \ \ \ \ \ \ \ \ \ \ \ \ \ \ \ \ \ \ \ \ \ \ \ \ \
\ \ \ \ \ \ \ \ M. M. Chowdhury\medskip \medskip

\textbf{Abstract}: We show that two new key exchange protocols with security
based on the triple decomposition problem may have security based on the
MSCSP.\medskip

\bigskip \textbf{1 Introduction}

Recently a new key exchange\ primitive based the triple DP (decomposition
problem) is proposed in [1] and the triple DP is defined in [1] as finding
the decomposition of a given element into three elements (that are not
known).

One purpose of inventing the above scheme of [1] is its security is based on
hard problems in braid groups such that a linear algebraic attack is not
possible. It is claimed in [1] that the security of the new scheme is based
on the triple DP in any $G$. If $G$ is a group / the private keys are
invertible then we show that the scheme in [1] is based on the CSP
(conjugacy search problem) or the MSCSP (multiple simultaneous CSP) hence
the algorithms in [1] are no more secure than using other key agreement
algorithms using the CSP or MSCSP and hence the new scheme in [1] can be
attacked feasibly using linear algebra if using braid groups (or the new
scheme can be attacked with any algorithm that gives solutions of the CSP or
MSCSP). There is a linear algebraic method to find solutions of the MSCSP
which has been used to attack the braid key exchange protocol of
Anshel-Anshel-Goldfeld [2] and this attack can be used to attack the new
scheme with linear algebra.\medskip

\textbf{2 Description of the New Protocols Based on the Triple Decomposition
Problem\medskip }

In this section the protocols are described using original portions of
sections 2 and 3 taken from the preprint [1] (hence the protocols are
described as exactly as in [1]).

\FRAME{ftbpF}{5.2797in}{5.3904in}{0pt}{}{}{j8qixa08.jpg}{\special{language
"Scientific Word";type "GRAPHIC";display "USEDEF";valid_file "F";width
5.2797in;height 5.3904in;depth 0pt;original-width 13.3337in;original-height
8.3333in;cropleft "0";croptop "1";cropright "1";cropbottom "0";filename
'C:/J8QIXA08.JPG';file-properties "XNPEU";}}

\bigskip \FRAME{ftbpF}{5.0505in}{1.542in}{0pt}{}{}{j8qixa09.jpg}{\special%
{language "Scientific Word";type "GRAPHIC";display "USEDEF";valid_file
"F";width 5.0505in;height 1.542in;depth 0pt;original-width
13.3337in;original-height 8.3333in;cropleft "0";croptop "1";cropright
"1";cropbottom "0";filename 'C:/J8QIXA09.JPG';file-properties "XNPEU";}}

\bigskip

\bigskip \FRAME{ftbpF}{5.0505in}{1.625in}{0pt}{}{}{j8qixa0a.jpg}{\special%
{language "Scientific Word";type "GRAPHIC";display "USEDEF";valid_file
"F";width 5.0505in;height 1.625in;depth 0pt;original-width
13.3337in;original-height 8.3333in;cropleft "0";croptop "1";cropright
"1";cropbottom "0";filename 'C:/J8QIXA0A.JPG';file-properties "XNPEU";}}

\bigskip

\FRAME{ftbpF}{5.0505in}{3.1669in}{0pt}{}{}{j8qixa0b.jpg}{\special{language
"Scientific Word";type "GRAPHIC";maintain-aspect-ratio TRUE;display
"USEDEF";valid_file "F";width 5.0505in;height 3.1669in;depth
0pt;original-width 13.3337in;original-height 8.3333in;cropleft "0";croptop
"1";cropright "1";cropbottom "0";filename 'C:/J8QIXA0B.JPG';file-properties
"XNPEU";}}

\textbf{2.1 Suggested Subgroup Parameters}

In this section the protocols are described using original portions of
sections 5 taken from the preprint [1] (hence the parameters are described
as exactly as in [1]).

\FRAME{ftbpF}{4.5878in}{2.2926in}{0pt}{}{}{j8qixa10.jpg}{\special{language
"Scientific Word";type "GRAPHIC";display "USEDEF";valid_file "F";width
4.5878in;height 2.2926in;depth 0pt;original-width 10.344in;original-height
4.1355in;cropleft "0";croptop "1";cropright "1";cropbottom "0";filename
'C:/Documents and Settings/fujitsu
pro/Desktop/preprint2006/J8QIXA10.JPG';file-properties "XNPEU";}}

\textbf{3 Security Of the Protocols based on the Triple Decomposition
Problem\medskip }

If $G$ is a group (it is suggested in [1] that $G$ may be a group an example
given in [1] for $G$ is the braid group) or the private keys are invertible
then we can show the following. In this section we give our new result that
the security of the new protocols in [1] is based on a system of equations
(1 \& 2 below), MSCSP or the CSP in $G$.\medskip

\textbf{3.1 The First Protocol\medskip }

Compute $%
O_{1}=pqr=(b_{1}y_{1})(y_{1}^{-1}b_{2}y_{2})(y_{2}^{-1}b_{3})=b_{1}b_{2}b_{3} 
$

Compute $%
O_{1}^{-1}pqJ_{I}r=O_{1}^{-1}(b_{1}y_{1})(y_{1}^{-1}b_{2}y_{2})J_{I}(y_{2}^{-1}b_{3}) 
$

$=(b_{3}^{-1}b_{2}^{-1}b_{1}^{-1})b_{1}b_{2}J_{I}b_{3}$

$=b_{3}^{-1}J_{I}b_{3}$ for $1\leq I\leq K_{1}\ \ \ $

\begin{equation}
b_{3}^{-1}J_{I}b_{3}\text{ for }1\leq I\leq K_{1}  \tag{1}
\end{equation}
For some integer $K_{1}$ and $J_{I}$ ($J_{I}$ may be braids) chosen by the
attacker.

Compute $%
pT_{I}qrO_{1}^{-1}=(b_{1}y_{1})T_{I}(y_{1}^{-1}b_{2}y_{2})(y_{2}^{-1}b_{3})O_{1}^{-1} 
$

$=b_{1}T_{I}b_{2}b_{3}(b_{3}^{-1}b_{2}^{-1}b_{1}^{-1})=b_{1}T_{I}b_{1}^{-1}$
for $1\leq I\leq K_{2}$

\begin{equation}
b_{1}T_{I}b_{1}^{-1}\text{ for }1\leq I\leq K_{2}  \tag{2}
\end{equation}
For some integer $K_{2}$ and $T_{I}$ chosen by the attacker. Observe the
elements $J_{I}$ are chosen from the $A_{3}$ (because of the commutativity
conditions of the protocols) and the elements $T_{I}$ are chosen from $A_{2}$%
. To find $b_{2}$ compute $b_{2}=b_{1}^{-1}O_{1}b_{3}^{-1}$ and now Bob's
private key is known and so the secret shared key can be constructed. Hence
from the systems of equations 1 and 2 the security of the protocol can be
based on the MSCSP [3] (which includes the CSP) hence we have shown that the
security of the new protocol in [1] is based on solving the MSCSP twice. A
very similar derivation show the security of the new protocol is also based
on two MSCSP with the unknowns $a_{1}$ and $a_{3}$. An observation from the
above is for any $G$ the above the security of the protocol can also be
based on (MSDSP)\ multiple simultaneous decomposition search problem for
example (using the above computations) by solving the equations $%
b_{1}b_{2}J_{I}b_{3}$ for $b_{3},b_{1}T_{I}b_{2}b_{3}$ for $b_{1}$ and then
solving for $b_{2}$ using $b_{1}$ and $b_{3}$ and using the publicly known
information (again there is a similar result using Alice's private keys) and
not the triple decomposition problem. Observe that for the possible specific
parameters suggested in [1] satisfy commutativity conditions such as $B_{2}$
commutes with $A_{2}$ etc. in addition to the required commutativity
conditions which are necessary for the protocol to work. We can use these
above additional commutativity conditions to show the security can be based
on the CSP as follows. We can solve MSCSP for $a_{1}$and $b_{3}$ as
described as above. Let $O_{2}=uvw=a_{1}a_{2}a_{3}$.

To recover the common secret key compute

$%
O_{2}^{-1}a_{1}(O_{1}b_{3}^{-1})a_{1}^{-1}O_{2}=a_{3}^{-1}a_{2}^{-1}(b_{1}b_{2})a_{2}a_{3}=a_{3}^{-1}(b_{1}b_{2})a_{3}, 
$

\begin{equation}
a_{3}^{-1}(b_{1}b_{2})a_{3}  \tag{3}
\end{equation}
similarly

$O_{1}b_{3}^{-1}(a_{1}^{-1}O_{2})b_{3}O_{1}^{-1}=$ $%
b_{1}(a_{2}a_{3})b_{1}^{-1}$

\begin{equation}
b_{1}(a_{2}a_{3})b_{1}^{-1}  \tag{4}
\end{equation}
we can solve for above $a_{3},b_{1}$ by solving the CSP with $%
b_{1}b_{2},a_{2}a_{3}$ Or we can solve the 5,6 below for $b_{1}$ and $a_{3}$
as follows (so again the protocol can be based on the MSCSP).

Attacker selects $V_{I}$ commuting with $a_{2}$ but not with $a_{3}$ or
select $V_{I}$ $\in B_{1}$

$%
O_{2}^{-1}a_{1}(V_{I})a_{1}^{-1}O_{2}=a_{3}^{-1}a_{2}^{-1}(V_{I})a_{2}a_{3}=a_{3}^{-1}V_{I}a_{3} 
$

\begin{equation}
a_{3}^{-1}V_{I}a_{3},\text{for }1\leq I\leq K_{3}  \tag{5}
\end{equation}
similarly the attacker selects $W_{I}$ commuting with $b_{2}$ but not with $%
b_{1}$ or select $V_{I}$ $\in A_{3}$

$O_{2}b_{3}^{-1}(W_{I})b_{3}O_{2}^{-1}=$ $b_{1}W_{I}b_{1}^{-1}$

\begin{equation}
b_{1}W_{I}b_{1}^{-1}\text{ for }1\leq I\leq K_{4}  \tag{6}
\end{equation}
The above result also holds when different subgroups are used that satisfy
the additional commutativity conditions (described above) for an arbitrary
G.\medskip

Observe that computing $b_{1}$ and $b_{3}$ from the MSCSP or CSP gives

$y_{1}=b_{1}^{-1}p$, $y_{2}^{-1}=rb_{3}^{-1},$ hence $%
b_{2}=(b_{1}^{-1}p)q(rb_{3}^{-1})$ ($a_{2}$ can be computed in a very
similar way).\medskip

To defend against the attack in section 3.1 of reconstructing the secret
shared key by solving the MSCSP the private keys of Alice, Bob are chosen so
that they not invertible.\medskip

\textbf{3.2 The Second Protocol\medskip }

The derivation to show the second protocol can be based on the MSCSP is
identical to the derivation for the first protocol except the elements $%
J_{I} $ are chosen from $S_{y_{2}}$, the elements $T_{I}$ are chosen from $%
S_{y_{1}}$ etc. hence the above observations also applies to the second
protocol. To defend against the attack in section 3.1 of reconstructing the
secret shared key by solving the MSCSP all the elements in the private keys
of Alice, Bob are chosen so that they are all not invertible.

\textbf{4 Conclusion\medskip }

We have shown that two new key exchange protocols with security based on the
triple decomposition problem may have security based on the MSCSP or the
MSDSP.\medskip

\medskip \textbf{References}

[1] A New Key Exchange Primitive Based on the Triple Decomposition Problem,
Yesem Kurt, Cryptology eprint archive,http://eprint.iacr.org/2006/378

[2] A Linear Algebraic Attack on the AAFG1 Braid Group Cryptosystem, 7th
Australasian Conference on Information Security and Privacy-2002, LNCS 2384,
Springer Verlag, pp. 176-189, 2002

[3] Ki Hyoung Ko, Tutorial on Braid Cryptosystems 3, PKC
2001,www.ipkc.org/pre\_conf/pkc2001/PKCtp\_ko.ps

[19] A New Key Exchange Protocol Based on the Decomposition Problem, V.
Shilparin and A. Ushakov, http://eprint.iacr.org/2005/447

\bigskip

\underline{Appendix}\medskip

We sketch the proof for the attacks considering the suggestion of 5.2.2 in
[1]. We recover Bob's private key as follows.

$%
s_{1}b_{1}s_{1}^{-1}(s_{2}y_{1}s_{2}^{-1})(s_{2}y_{1}^{-1}s_{2}^{-1})(s_{3}b_{2}s_{3}^{-1})(s_{4}y_{2}^{-1}s_{4}^{-1})(s_{4}y_{2}s_{4}^{-1})(b_{3})=
$

$\allowbreak s_{1}b_{1}s_{1}^{-1}s_{3}b_{2}s_{3}^{-1}b_{3}=O_{1}$

$%
O_{1}^{-1}s_{1}b_{1}s_{1}^{-1}(s_{2}y_{1}s_{2}^{-1})(s_{2}y_{1}^{-1}s_{2}^{-1})(s_{3}b_{2}s_{3}^{-1})(s_{4}y_{2}s_{4}^{-1})(s_{4}H_{I}s_{4}^{-1})
$

$(s_{4}y_{2}^{-1}s_{4}^{-1})(b_{3})=\allowbreak
O_{1}^{-1}s_{1}b_{1}s_{1}^{-1}s_{3}b_{2}s_{3}^{-1}s_{4}Hs_{4}^{-1}b_{3}=$ $%
b_{3}^{-1}s_{4}H_{I}s_{4}^{-1}b_{3}$.

Hence $b_{3}$ can be found by solving the MSCSP.

Then $y_{2}=(s_{4}^{-1}rb_{3}^{-1}s_{4})^{-1}$

Now select $J_{I}$ form $A_{2}$.

$%
s_{1}b_{1}s_{1}^{-1}(s_{2}y_{1}s_{2}^{-1})(s_{2}J_{I}s_{2}^{-1})(s_{2}y_{1}^{-1}s_{2}^{-1})(s_{3}b_{2}s_{3}^{-1})(s_{4}y_{2}s_{4}^{-1})(s_{4}y_{2}^{-1}
$

$s_{4}^{-1})(b_{3})O_{1}^{-1}=$

$%
s_{1}b_{1}s_{1}^{-1}(s_{2}y_{1}s_{2}^{-1})(s_{2}J_{I}s_{2}^{-1})(s_{2}y_{1}^{-1}s_{2}^{-1})(s_{3}b_{2}s_{3}^{-1})(s_{4}y_{2}s_{4}^{-1})(s_{4}y_{2}^{-1}s_{4}^{-1})
$

$%
(b_{3})((s_{1}b_{1}s_{1}^{-1})(s_{2}y_{1}s_{2}^{-1})(s_{2}y_{1}^{-1}s_{2}^{-1})(s_{3}b_{2}s_{3}^{-1})(s_{4}y_{2}^{-1}s_{4}^{-1})(s_{4}y_{2}s_{4}^{-1})(b_{3}))^{-1}=
$

$%
=s_{1}b_{1}s_{1}^{-1}(s_{2}y_{1}s_{2}^{-1})(s_{2}J_{I}s_{2}^{-1})(s_{2}y_{1}^{-1}s_{2}^{-1})s_{1}b_{1}^{-1}s_{1}^{-1} 
$

$=s_{1}b_{1}s_{1}^{-1}(s_{2}J_{I}s_{2}^{-1})s_{1}b_{1}^{-1}s_{1}^{-1}$

Hence $b_{1}$ can be found by solving the MSCSP.

Then $y_{1}=s_{1}b_{1}^{-1}s_{1}^{-1}s_{2}^{-1}ps_{2}$

Then we can recover Bob's second private key as

$(s_{2}y_{1}s_{2}^{-1})q(s_{4}y_{2}s_{4}^{-1})=(s_{3}b_{2}s_{3}^{-1})$

Now we have Bob's private key, the shared key is recovered as

$%
a_{1}(s_{1}x_{1}s_{1}^{-1})(s_{1}b_{1}s_{1}^{-1})(s_{1}x_{1}^{-1}s_{1}^{-1})(s_{2}a_{2}s_{2}^{-1})(s_{3}x_{2}s_{3}^{-1})(s_{3}b_{2}s_{3}^{-1})(s_{3}x_{2}^{-1}s_{3}^{-1})
$

$(s_{4}a_{3}s_{4}^{-1})b_{3}=$

$%
a_{1}(s_{1}x_{1}b_{1}x_{1}^{-1}s_{1}^{-1})(s_{2}a_{2}s_{2}^{-1})(s_{3}x_{2}b_{2}x_{2}^{-1}s_{3}^{-1})(s_{4}a_{3}s_{4}^{-1})b_{3}=shared 
$ $key$

There are similar attacks for each of our above attacks.

\end{document}